\documentclass[prd,preprint,superscriptaddress]{revtex4}

\newcommand{\be}{\begin{equation}}
\newcommand{\ee}{\end{equation}}
\newcommand{\ba}{\begin{eqnarray}}
\newcommand{\ea}{\end{eqnarray}}
\newcommand{\no}{\noindent}
\newcommand{\n}{\label}

\begin{document}

\title{Symmetries leading to inflation}

\author{Juan M. Aguirregabiria}
\affiliation{F\'{\i}sika Teorikoa eta Zientziaren Historia Saila,
Zientzi Fakultatea,
Euskal Herriko Unibertsitatea,
644 Posta Kutxatila, 48080 Bilbao, Spain}
\author{Luis P. Chimento}
\affiliation{Departamento de F\'{\i}sica,
Facultad de Ciencias Exactas y Naturales,
Universidad de Buenos Aires,
Ciudad  Universitaria,  Pabell\'on  I,
1428 Buenos Aires, Argentina.}
\author{Alejandro S. Jakubi}
\affiliation{Departamento de F\'{\i}sica,
Facultad de Ciencias Exactas y Naturales,
Universidad de Buenos Aires,
Ciudad  Universitaria,  Pabell\'on  I,
1428 Buenos Aires, Argentina.}
\author{Ruth Lazkoz}
\affiliation{F\'{\i}sika Teorikoa eta Zientziaren Historia Saila,
Zientzi Fakultatea,
Euskal Herriko Unibertsitatea,
644 Posta Kutxatila, 48080 Bilbao, Spain}

\date{\today}

\begin{abstract}

We present here the general transformation that leaves unchanged the form of
the field equations for perfect fluid Friedmann--Robertson--Walker and
Bianchi V cosmologies. The symmetries found can be used as algorithms for
generating new cosmological models from existing ones. A particular case of
the general transformation is used to illustrate  the crucial role played by the number of scalar fields in the
occurrence of inflation. Related to this, we also study the existence and
stability of Bianchi V power law solutions.

\end{abstract}

\maketitle

\section{Introduction}

Symmetries are the cornerstone of the development of  modern physics and in
particular of the description of the Universe at large scales. Even General
Relativity, the framework used for that description, has its roots in symmetry
requirements. Not long ago, a new proposal for exploiting  the symmetries of
the Einstein field equations was  made \cite{si}. In this framework two
spacetimes are said to be equivalent if the corresponding set of equations is
form invariant under the action of a given transformation. This unusual
concept of equivalence has been successfully applied to the study of
equivalences among different cosmological models, and in particular to those
exhibiting inflation.

The equations  that govern the evolution of spatially flat
Friedmann--Robertson--Walker (FRW) cosmologies filled with a perfect fluid or
massive scalar fields happen to admit a peculiar type of form invariance. It
directly relates an increase in the energy density of the model with an
increase in its expansion rate. This provides a mechanism for using
non-accelerating FRW models as seeds for inflationary cosmologies.

Remarkably, this link between the energy density and the expansion rate is the
key feature of the assisted inflation proposal. According to it, for some
cosmological models, the occurrence of inflation is directly related to the
number of scalar fields driving the expansion \cite{lms}. This cooperative
effect can be easily  illustrated in terms of the form invariance we alluded
to. In particular, for the expansion rate of a spatially flat FRW model to
increase by a factor of $n$ it is only necessary that  the energy density gets
multiplied by $n ^2$. In the language of scalar fields this just means
that a universe containing a single selfinteracting field has 
transformed into one with $n$ fields interacting with themselves but not
amongst them.

The natural question that comes to mind is whether this form invariance
symmetry is just a very special feature of spatially flat FRW, rather than
commonplace. We have addressed here this question, and our results show that
form invariance transformations do exist for any FRW models and their simplest
generalization,  Bianchi V models.

In sections I and II we  outline the details of the transformations of the
metric functions, pressure and energy density of perfect fluid FRW and Bianchi
V cosmologies leading to unchanged field equations. Then, we exploit the
equivalence between that kind of matter content and a scalar field with a
self-interaction potential to write equivalent transformation  rules for them.
The existence of this symmetry provides us with algorithms to generate new
cosmological solutions from known ones. We are mainly concerned here by the
possibility of generating inflationary solutions from others that do not show
that behaviour.

A straightforward particularization of the transformation allows obtaining
spacetimes with a Hubble factor that is a constant times the original one. As
a consequence, the transformed deceleration factor becomes smaller than the
original one, provided this constant is larger than unity. One key feature
here is that this information on the spacetime expansion is obtained without
knowing the metric functions explicitly.

This is, as we see, very similar to what happened in the assisted inflation
scenario (we will dwell on the identification of the analogies in the sections
below). Previous studies on this topic have focused  on power-law solutions
because  they are late time attractors for the evolution of both FRW
\cite{lms} and Bianchi I-VI$_{\mbox 0}$ \cite{accz} models (see as well
\cite{Gre} for a discussion regarding general geometries). We reach in Section
III an analogous result in the Bianchi V case, and in particular, we will see
that the expression of the potential in terms of the scale factor for the
power-law solutions gives the clue to a method of obtaining new exact scalar
field solutions up to quadratures.

The discussion fits into a method to obtain solutions that, as will be
explained in Section IV, has turn out to be fruitful before. The scalar field
potentials, or the equation of state in the case of perfect fluids, and the
evolution of the scale factor are derived from the history of the potential.
We use these solutions to present simple examples of the action of the
transformation on models with well known potentials.
Finally, we outline our future prospects and main conclusions in section V.

\section{Form invariance symmetry in FRW spacetimes}

Let us consider a FRW spacetime (isotropic and spatially homogeneous)
with curvature
$k$, which is described by the metric
\begin{equation}
\label{rw}
ds^{2}=-dt^{2}+a^{2}(t)\left[(1-kr^2)^{-1}\,dr^{2}+
r ^2\left(d\theta^{2}+\sin ^2\theta\, d\phi^{2}\right)\right].
\end{equation}
If the source of the geometry is a perfect fluid with 
energy density $\rho$ and identical pressure
 $p$ along the three spatial directions (isotropic
perfect fluid), then the model is governed by the Friedmann equation
\begin{equation}
\label{00F}
3H^{2}+3\frac{k}{a^2}=\rho
\end{equation} 
and the energy conservation equation 
\be
\n{co}
\dot\rho+3H(\rho+p)=0,
\ee
where, $H=\dot a/a$ is the Hubble factor, as usual. Given
a different perfect fluid with energy density $\bar\rho$ and pressure $\bar p$,
the corresponding equations take the form
\begin{equation}
\label{00b}
3\bar H^{2}+3\frac{k}{\bar a^2}=\bar\rho,
\end{equation} 
\be
\n{cob}
\dot{\bar\rho}+3\bar H(\bar\rho+\bar p)=0.
\ee
Our goal is to obtain a transformation that leaves the form of the 
system of equations (\ref{00b})--(\ref{cob}) unchanged. In other words,
we want to find the symmetry transformation that maps
(\ref{00b})--(\ref{cob}) into (\ref{00F})--(\ref{co}). Since we 
mean to obtain this transformation explicitly,  we make the
following ansatz:
\ba
\n{ta}
\bar a&=&\bar a(a,H,\rho,p),\\
\n{th}
\bar H&=&\bar H(a,H,\rho,p),\\
\n{tr}
\bar\rho&=&\bar\rho(a,H,\rho,p),\\
\n{tp}
\bar p&=&\bar p(a,H,\rho,p).
\ea 
{F}rom equations (\ref{00F})--(\ref{co}) and
(\ref{00b})--(\ref{cob}) we obtain the Raychaudhuri equation

\be
\n{Hp}
\dot H=-\frac{1}{2}(\rho+p)+\frac{k}{a^2}
\ee
and its transformed version
\be
\n{Hpb}
\dot{\bar H}=-\frac{1}{2}(\bar\rho+\bar p)+\frac{k}{\bar a^2}.
\ee

In what follows, we are going to use the fact that a transformation can be
regarded as a symmetry transformation when it does not impose restrictions on
the functions appearing in the equations, that is, when one always obtains
identities for any function. If we differentiate Eq. (\ref{th}) a term
proportional to $\dot p$ arises, but if we insert the expression in the
l.h.s.~of Eq. (\ref{Hpb}), it can be noticed immediately that it must vanish
identically for there are no terms proportional to $\dot p$ in the  r.h.s.
Then, the first conclusion we draw is that $\bar H=\bar H(a,H,\rho)$. If we
replace now Eqs. (\ref{ta}) and (\ref{tr}) in Eqs. (\ref{00b}) and (\ref{cob}), and use
the same argument we conclude that $\bar a=\bar a(a,H,\rho)$ and
$\bar\rho=\bar\rho(a,H,\rho)$. 

The next step is to calculate $\bar H$  from Eq. (\ref{ta})  using the
definition $\bar H=\dot {\bar a}/\bar a$, so that
\be
\n{bh}
\bar H=\frac{a}{\bar a}\frac{\partial\bar a}{\partial a}H+\frac{1}{\bar a}
\frac{\partial\bar a}{\partial H}
\left[\frac{k}{a^2}-\frac{1}{2}(\rho+p)\right]-\frac{3H(\rho+p)}{\bar a}
\frac{\partial\bar a}{\partial \rho},
\ee
where equations (\ref{co}) and (\ref{Hp}) have been used. Since $\bar H$
does not depend on $p$,  the coefficient of $p$ in Eq. (\ref{bh}) must vanish
as well. For that reason,
\be
\n{cp}
\frac{\partial\bar a}{\partial \left(3H^2\right)}+
\frac{\partial\bar a}{\partial\rho}=0,
\ee
and the  general solution to the latter  is 
$\bar a=\bar a\left(a,\rho-3H^2\right)=
\bar a\left(a,3k/a^2\right)$, that
is,  $\bar a$ depends on $a$ only.

Summarizing, the transformation turns out
to be
\be
\n{taf}
\bar a=\bar a(a),
\ee
\be
\n{thf}
\bar H=\frac{\partial\ln \bar a}{\partial\ln a}\, H,
\ee
\be
\n{trf}
\bar\rho=3\left[\frac{\partial\ln \bar a}{\partial\ln a}\right]^2H^2+
\frac{3k}{\bar a^2},
\ee
\be
\n{tpf}
\bar p=-\bar\rho-2\dot{\bar H}+2\frac{k}{\bar a^2},
\ee
where $\dot{\bar H}$ has to be calculated using Eq. (\ref{thf}).
Here, $\bar a=\bar
a(a)$ is the only ``parameter'' of the symmetry transformation.
Once more, we must emphasize  the
general character of the transformation as no
 assumption on the equation of  state of the
fluid had to be made.

Now, it is interesting to investigate the transformation properties of 
other physical parameters, so that one can deepen in the comparison
between the features of the two models. For instance, 
the deceleration parameter
\begin{equation} \label{q}
q (t)=-\frac{\ddot{a}}{aH^2}
\end{equation}
%
transforms as
\be
\n{tq}
\bar q+1=\left[\frac{\partial\ln \bar a}{\partial\ln a}\right]^{-1}(q+1)+
\frac{\partial}{\partial\ln a}
\left[\frac{\partial\ln \bar a}{\partial\ln a}\right]^{-1}
\ee
under the symmetry transformations (\ref{tr})--(\ref{tp}).

As an example let us look at the power--law transformation $\bar a=a^n$. In
this particular case, we obtain from Eqs. (\ref{thf}) (\ref{trf}) and
(\ref{tpf}) the transformation rules $\bar H=nH$,

\be
\n{br}
\bar\rho=n^2\left[\rho-3\frac{k}{a^2}\right]+3\frac{k}{ a^{2n}}
\ee
and
\be
\n{bp}
\bar p=-3n^2H^2-2n\dot H-\frac{k}{a^{2n}}.
\ee
The transformation rule for the deceleration parameter is
\be
\n{bq}
\bar q=-1+\frac{q+1}{n}.
\ee

\no It becomes negative for $n$ high enough, showing in this case that the
inflation is more likely. In other words, the strong energy condition (SEC)
$\bar\rho+3\bar p= -6(n^2H^2+n\dot H)>0$ is violated for large $n$. So, we
will be interested in the cases with $n>1$, where the transformed evolution
will always be closer to de Sitter spacetime than the original one.

The fluid interpretation of the scalar field has proven very useful in the
study of the inflationary and quintessence scenarios \cite{cjp}. The
energy-momentum tensor of the scalar field may be written in the perfect fluid
form
\begin{equation}\label{Tfluid}
T_{ik}= (p+\rho) u_{i}u_{k}+pg_{ik},
\ee
 if one defines
\begin{eqnarray}\label{rhophi}
\rho & = &  \frac{1}{2} \dot\phi^2 + V(\phi),
\\\label{rhophi2}
   p & = &  \frac{1}{2} \dot\phi^2 - V(\phi),
\end{eqnarray}
where we have taken into account that the scalar field $\phi$ depends
on $t$ only.
Inserting Eqs. (\ref{rhophi})--(\ref{rhophi2}) in
Eqs. (\ref{br})--(\ref{bp}) we find that
\be
\n{ff}
\dot{\bar\phi}^2=n\dot\phi^2+2k\left[\frac{1}{a^{2n}}-\frac{n}{a^2}\right],
\ee
\be
\n{vf}
\bar V=nV+3n(n-1)H^2+2k\left[\frac{1}{a^{2n}}-\frac{n}{a^2}\right].
\ee

Asymptotically (i.e., when $a\to\infty$) and for large and positive $n$ the
transformation (\ref{br}) reduces  to $\bar\rho\approx n^2\rho$ and we recover
in this limit the results found in \cite{si} for the particular case of flat
FRW metrics.  Under these specific conditions one can speak in terms of
a  single field model transforming into a multifield model in the
fashion of assisted inflation.

\section{Bianchi V models}

We wish to investigate now the form invariance symmetry
of the Einstein equations of an anisotropic universe described by
the Bianchi V metric. These spacetimes are the simplest generalizations
of FRW open universes and they possess a non-abelian group
of three isometries. We write the line element as  (cf.~\cite{Jos})
\begin{equation}
 d s^2 = -d t^2+e ^{f(t)}d z^2 + G(t) e^{z}
 \left(e^{h(t)}\, d x^2 +e^{-h(t)}\,d y^2\right),
 \label{metric}
\end{equation}
with $G(t)\ge 0$. Considering once again a universe filled with an isotropic
perfect fluid, the Einstein equations for~(\ref{metric}) are
\begin{eqnarray}
\label{einstein_1}
&&\frac{\ddot G}{G}+\frac{{\dot G}}{2G}{\dot f}+p-\rho-e ^{-f}=0,\\
\label{einstein_2}
&&\ddot h+\dot h\left(\frac{\dot f}{2}+\frac{\dot G}{G}\right)=0,\\
\label{einstein_3}
&&\frac{\dot G}{G}-\dot f=0,\\
\label{einstein_4}
&&\frac{\ddot G}{G}-\frac{\dot G \dot f}{2G}-\frac{\dot G ^2}{2G ^2}+
\frac{\dot h ^2}{2}+p+\rho+\frac{e ^{-f}}{2}=0,\\
\label{conservation}
&&\dot\rho +\frac{3 \dot G}{2 G}(\rho+p)=0.
\end{eqnarray}
{F}rom Eqs. (\ref{einstein_2}) and (\ref{einstein_3}) one readily gets
\begin{eqnarray}
\label{einstein_5}
&&e ^f=G,\\
\label{einstein_6}
&&\dot h=\frac{A}{G^{3/2}},
\end{eqnarray}
with the  integration constant $A\neq 0$. Combining the remaining
Einstein equations in terms of the new variables
\be
\n{nv}
a=G^{1/2},    \qquad   H=\frac{\dot a}{a}=\frac{\dot G}{2G},
\ee
we get
\begin{equation} 
3H^2=\rho+\frac{3}{4a^2}+\frac{A^2}{4a^6},
\label{combination}
\end{equation}
\begin{equation}
\dot\rho+3H(\rho+p)=0\label{con}.
\end{equation}

\no These equations are equivalent to those of negatively curved FRW models
filled with a perfect fluid and stiff matter. The shear's contribution to the
expansion is precisely the last term in Eq. (\ref{combination}) and basically can
be associated with a free scalar field. Equation (\ref{combination}) together
with the conservation equation (\ref{con}) forms the set of equations to be
solved.

\subsection{Form invariance symmetry}

Remarkably, the pair formed by equations (\ref{combination}) and (\ref{con})
is form-invariant as well. This means that for a different fluid with density
$\bar\rho$ and pressure $\bar p$ the equations
\begin{eqnarray}
&&3\bar H^2=\bar\rho+\frac{3}{4\bar a^2}+\frac{\bar A^2}{4\bar a^6},\\
&&\dot{\bar\rho} +3\bar H(\bar \rho+\bar p)=0
\end{eqnarray}
become equations (\ref{combination})--(\ref{con}) under the symmetry
transformation
\be
\n{tabf}
\bar a=\bar a(a),
\ee
\be
\n{thbf}
\bar H=\frac{\partial\ln \bar a}{\partial\ln a}\, H,
\ee
\be
\n{trbf}
\bar\rho=3\left[\frac{\partial\ln \bar a}{\partial\ln a}\right]^2H^2-
\frac{3}{4\bar a^2}-\frac{\bar A^2}{4\bar a^6},
\ee
\be
\n{tpbf}
\bar p=-\bar\rho-2\dot{\bar H}-\frac{1}{2\bar a^2}-\frac{\bar A^2}{2\bar a^6},
\ee
where $\dot{\bar H}$ has to be calculated using Eq. (\ref{thbf}). Here,
$\bar a=\bar a(a)$ is the ``parameter'' of the transformation.

Since our main objective is to use this symmetries in the context of
inflation, it is of interest to exploit the customary equivalence between a
perfect fluid and a self-interacting scalar field with potential $V(\phi)$
when $\bar a=a^n$. 
 In this case equations (\ref{combination})--(\ref{con}) become
\be
\n{00}
3H^2=\frac{1}{2}\dot\phi^2+V+\frac{3}{4a^2}+\frac{A^2}{4a^6},
\ee
\be
\n{kg}
\ddot\phi+3H\dot\phi+\frac{\partial V}{\partial\phi}=0,
\ee
where the last one is the Klein-Gordon equation. 
Transforming Eqs. (\ref{rhophi})--(\ref{rhophi2}) according to
Eqs. (\ref{trbf})--(\ref{tpbf}) we obtain the transformation properties of the
scalar field and the corresponding potential
\ba
\label{tphi}
&&\dot {\bar \phi} ^2=n\dot \phi ^2
+\frac{1}{2}\left(\frac{n}{a^2}-\frac{1}{a^{2n}}\right)
+\frac{1}{2}\left(\frac{nA^2}{a^6}-\frac{\bar A^2}{a^{6n}}\right),\\
&&\bar V=nV+3n(n-1)H^2+\frac{n}{2a^2}-\frac{1}{2a^{2n}}.
\label{tpo}
\ea

It can be shown that the transformation rule for the deceleration
factor under the transformation $\bar a=a^n$ is again Eq. (\ref{bq}).

\subsection{Asymptotic behaviour and power-law solutions}

The combined measurements of the cosmic microwave background  temperature
fluctuations and the distribution of galaxies on large scales seem to imply
that the Universe may  be flat or nearly flat \cite{OS95,Bahcall99,Efst}. The
equivalence between the equations (\ref{combination})--(\ref{con}) for the
Bianchi V model and those of open  FRW, permits to introduce the cosmological
density parameter $\Omega$, defined as the ratio of the scalar field energy
density $\rho$ to the asymptotic critical density $\rho_c=3H^{2}$. The
asymptotic critical density corresponds to the asymptotic flat spacetime 
solution of
equation (\ref{00}). Thus, inserting $\Omega\equiv\rho/\rho_c=\rho/3H^2$ in
Eqs. (\ref{combination})--(\ref{con}), we find the dynamical equation for the
cosmological density parameter:
\begin{equation}
\dot\Omega=\left[6\frac{a^4+A^2}{3a^4+A^2}-3\left(1+
\frac{p}{\rho}\right)\right]\,\Omega (1-\Omega)H.
\label{om2}
\end{equation}
Stable constant solutions of this equation are relevant to understand the
asymptotic behaviour of that spacetime, where Eq. (\ref{om2}) reduces to
\begin{equation}
\dot\Omega=-\left(1+3\frac{p}{\rho}\right)\,\Omega (1-\Omega)H
+O\left(\frac{1}{a^4}\right)\,.
\label{om2bis}
\end{equation}

The constant solution of Eqs. (\ref{om2}) and (\ref{om2bis}), compatible with
observations at late times, is $\Omega = 1$. It is asymptotically stable for
expanding universes ($H > 0$) when the SEC is violated. Hence the model
inflates, i.e., $\ddot{a}/a = -(\rho+3p)/6 > 0$ and the expansion of the
universe is driven by a gravitationally repulsive stress.

The general form of the perfect fluid Bianchi V solution,  up to a
quadrature, has been known \cite{Ellis} for long (see also (\cite{Heck},
\cite{Ruban},\cite{EllMac}). Here, however, we adopt a different perspective 
in what exact
solutions are regarded. In order to investigate the existence of power-law
solutions in the Bianchi V model with a self-interacting scalar field and
their stability, we combine the scaling parameter $\omega=-2\dot H/3H^2$ with
equations (\ref{00}) and (\ref{kg}), so that we get

\be
\n{do}
\dot\omega=(\omega-2)\left[\frac{\dot
V-({H}/{a^2})}{V+({1}/{2a^2})}+3H\omega\right].
\ee
The fixed point solution of Eq. (\ref{do}), $\omega=\omega_0=\mbox{constant}$,
corresponds to
\ba
a&=&t^{{2}/{3\omega_0}},\label{vdot0}\\
V&=&\frac{2\left(2-\omega_0\right)}
{3\omega_0^2a^{3\omega_0}}-\frac{1}{2a^2},\label{vdot}
\ea
where the scalar field can be calculated from Eq. (\ref{00}):
\be
\frac{1}{2}\dot{\phi}^2=\frac{2}{3\omega_0 t^2}-\frac{1}{4t^{4/3\omega_0}}
-\frac{A^2}{4t^{4/\omega_0}}.\label{vdot1}
\ee
Furthermore, equation (\ref{do}) becomes
\be
\dot\omega=3(\omega-2)(\omega-\omega_0)H.
\ee
Therefore, for the potential (\ref{vdot}) the power-law solution is
asymptotically stable whenever be $\omega_0<2$. When $\omega_0<2/3$ the SEC is
violated and there is an accelerated scenario. If $2/3<\omega_0<2$ we can 
construct an inflationary model
by substituting $\omega_0/n$ for $\omega_0$
and taking $n$ large enough, because 
\be
\n{aex}
\bar a=t^{2n/3\omega_0}
\ee
is an  asymptotically stable exact power-law solution
of Eqs. (\ref{00})--(\ref{kg}) for
the potential
\be
\n{pex}
\bar V=\left(2-\frac{\omega_0}{n}\right)\frac{2n^2}{3\omega_0^2
\bar a^{3\omega_0/n}}-\frac{1}{2\bar a^2},
\ee
where the scalar field can be calculated from Eq. (\ref{00}):
\be
\n{cex}
\frac{1}{2}\dot{\bar\phi}^2=\frac{2n}{3\omega_0 t^2}-
\frac{1}{4t^{4n/3\omega_0}}
-\frac{A^2}{4t^{4n/\omega_0}}.
\ee
Quantities $(\dot\phi,V)$ and $(\dot{\bar\phi},\bar V)$
are related by transformations (\ref{tphi})--(\ref{tpo}). Also from
Eqs. (\ref{tabf}), (\ref{thbf}) and (\ref{aex}), we obtain
\be
\bar a=a^n,    \qquad      \bar H=nH.
\ee

For large $n$ the energy density of the scalar field transforms as
$\bar\rho\approx n^2\rho$ showing that the cumulative effects of
adding energy density in the 00-component of the Einstein equations leads to an
accelerated scenario. In this regime $\bar\phi\approx 2(n/3\omega_0)^{1/2}\ln
t$ and the potential is exponential
\be
\n{poa}
\bar V(\phi)\approx \frac{2n^2}{3\omega_0^2}\left(2-
\frac{\omega_0}{n}\right){\mbox e}^{-\sqrt{3\omega_0/n}\,\phi}.
\ee
When $n$ is an integer it could represent the number of scalar fields
contained in the Bianchi V universe. This particular multifield Bianchi V
problem is equivalent to realize the usual assisted inflation in the FRW model
with $n$ identical non interacting scalar fields, in a negatively curved
space-time filled with an extra free scalar field. Hence we have extended the
previous results obtained in \cite{si} for FRW to the Bianchi V type metrics.
Again we have seen that form invariance of the Einstein equations leads to a
simple generalization of the assisted inflation linking two different
cosmological models, one of which, is accelerated.

\section{Transformation between closed form solutions}

We seek now models for which the potential $V(\phi)$ and the scale factor
evolution $a(t)$ can be obtained in closed form and exhibit the action of the
power--law transformation in simple terms. The expression (\ref{pex}) for the
potential suggests to start looking for exact solutions where the scale factor
is the independent variable. In Ref.~\cite{scal} it was shown the reduction of
the system (\ref{00F})(\ref{co}) to quadratures using as input the history of
the potential. In this case Eq. (\ref{co}) becomes (\ref{kg}), we write the
potential as
\begin{equation}
\label{Va}
V[\phi (a)] = {\frac{F(a)}{a^{6}}}
\end{equation}
and make the change of
variables $dt=a^3 d\eta$ in Eq. (\ref{kg}),
\begin{equation} \label{KG1}
\frac{d^2\phi}{d\eta^2}+a^6 \frac{dV}{d\phi}=0,
\end{equation}
so that we obtain the first integral
\begin{equation}
\label{KG2}
\frac {1}{2}{\dot \phi }^2+V(\phi)-\frac 6{a^6}\, 
\int da\frac{F}a=\frac C{a^6}, 
\end{equation}
where $C$ is an arbitrary integration constant.
Then, using equations
(\ref{00F}) and (\ref{kg}) we obtain
\begin{equation}
\Delta t = {\sqrt{3}\int \frac{da}{a}\left[
{\frac 6{a^6}\int }da{\frac Fa}+{\frac C{a^6}}-3{
\frac k{a^2}}\right]^{-1/2}},
\label{12} 
\end{equation}
\begin{equation}
\Delta \phi = {\sqrt{6}\int \frac{da}a\left[\frac{-F+6\int da F/a+C}
{6\int da F/a+C-3ka^4}\right]}^{1/2},
\label{13} 
\end{equation}
\noindent where $\Delta t\equiv t-t_0$, $\Delta \phi \equiv \phi -\phi _0$
and $t_0$, $\phi _0$ are two other arbitrary integration constants.

This unconventional procedure of solving the field equations starting from the
history of the potential was also used to obtain exact solutions in two and
four dimensional spacetimes with a scalar field and a perfect fluid in Refs.
\cite{scal,Cos}. An interesting simple class of models arise for
potentials with power--law histories $F=B a^m$ and $C=0$. In this case,
$\dot\rho_\phi=-3H\dot\phi^2<0$ implies $m<6$ for $H>0$, $\dot\phi^2>0$
implies $B/m>0$. Then, for positive definite potentials, hence $B>0$, we have
$0<m<6$. We find hyperbolic potentials for $m\neq 4$ and $k=1$
\begin{equation} \label{V1}
V(\phi)=\left(\frac{m}{2}\right)^{\frac{6-m}{4-m}}B^{2/(m-4)}
\left\{\cosh^2\left[\frac{(4-m)\Delta\phi}{2(6-m)^{1/2}}\right]\right\}
^{\frac{6-m}{4-m}}\,,
\end{equation}
\noindent
or $k=-1$
\begin{equation} \label{V-1}
V(\phi)=\left(\frac{m}{2}\right)^{\frac{6-m}{4-m}}B^{2/(m-4)}
\left\{\sinh^2\left[\frac{(4-m)\Delta\phi}{2(6-m)^{1/2}}\right]\right\}
^{\frac{6-m}{4-m}}\,,
\end{equation}
\noindent
while an exponential potential arise for $m=4$, $k=-1$ or $k=1$ with $B>2$,
\begin{equation} \label{V4}
V(\phi)=B\exp\left[-\left(2\frac{B-2k}{B}\right)^{1/2}\Delta\phi\right].
\end{equation}

As we see, potentials of this class include hyperbolic potentials that are
relevant to describe the scalar dark matter
\cite{Matos:1999et,Matos:2000ng,Matos:2000ss}; and they also include the
exponential potential, which is typically considered when modelling inflation,
and is motivated by supergravity theories \cite{cop,eas}.

For these models, the scale factor can be obtained in closed, implicit form
when $m\neq 4$ in terms of the hypergeometric function. For $k=1$ we obtain
\begin{equation} \label{30}
\Delta t=\left(\frac{2m}{B}\right)^{1/2} \frac{a^{3-m/2}}{6-m}\,{}_2F_1
\left(\frac{1}{2},\frac{m-6}{2(m-4)},\frac{3m-14}{2(m-4)},
\frac{ma^{4-m}}{2B}\right)\,,
\end{equation}
\noindent
while for $k=-1$ we get
\begin{equation} \label{38}
\Delta t=a \,{}_2F_1\left(\frac{1}{2},\frac{1}{m-4},\frac{m-3}{m-4},
-\frac{2B}{m}a^{m-4}\right).
\end{equation}
\noindent We wish to see the action of the symmetry transformation on this
class of models, linking non-accelerated expansions with accelerated ones. So
we start looking at their late time behavior.

For $k=1$ and $4<m<6$, the scale factor has a bounce and its asymptotic
behavior for large times is $t^{2/(6-m)}$; while for $0<m<4$, the scale factor
has an upper bound so that we will not consider this case any further. For
$k=-1$ and $0<m<4$ the scale factor evolves from a unaccelerated initial
phase with $a\simeq\Delta t^{2/(6-m)}$ at small times to a linear expansion at
late times; while for $4<m<6$ the scale factor evolves from a linear stage at
small times to an accelerated stage $a\simeq t^{2/(6-m)}$ at late times.
Finally, for $k=\pm 1$ and $m=4$, the expansion is linear.

Taking into account Eq.
(\ref{KG2}) we observe that the action of the symmetry transformation
(\ref{taf})--(\ref{tpf}) becomes the map $(F(a),C)\to(\bar F(\bar a),\bar
C)$ where Eq. (\ref{12}) remains invariant. That is, the set of the
solutions that can be expressed in terms of quadratures transforms into
itself. In the case of the power--law transformation (\ref{ff})--(\ref{vf})
we get
\begin{equation} \label{barFrw}
\bar F=na^{6(n-1)}\left[F+\left(n-1\right)\left(6\int\frac{da}{a}F+
C\right)\right]-
n\left(3n-1\right)ka^{6n-2}+2ka^{4n}.
\end{equation}
\noindent
For the models with power--law history
of their potential the transformation becomes
\begin{equation} \label{barFprw}
\bar F=nB\left(1+6\frac{n-1}{m}\right)a^{6(n-1)+m}-
n(3n-1)ka^{6n-2}+2ka^{4n}.
\end{equation}
\noindent So, the requirement that the transformed potential has a power--law
history, i.e. $\bar F=\bar B\bar a^{\bar m}$ can be satisfied for $k=-1$ with
$\bar B=n(3n-1)$ and $\bar m=(32-6m)/(6-m)=6-2/n$. The requirements that $m>0$
and $\bar m>0$ constraint the transformation exponent to $1/3<n<3$ and the
history exponents to the interval $(0,16/3)$. We observe that the
transformation with $n>1$ maps unaccelerated models with $m<4$ into
accelerated ones with $\bar m>4$, while the linearly expanding model with
$m=4$ transforms into itself.

As a second example we consider transformations for $k=\pm 1$ that allow an
arbitrary large exponent. In this case we will request that the transformed
potential has power--law history only asymptotically for $n\gg 1$ and large
scale factor. This case occurs for $m=4$ and we have
\begin{equation} \label{barB}
\bar B=n\left\{B\left[1+\frac{3}{2}(n-1)\right]-(3n-1)k\right\}
\end{equation}
\noindent and again $\bar m=6-2/n$. This transformation maps the linearly
expanding model with exponential potential into a model with deceleration
parameter so close to $-1$ as required while $n$ is made large enough.
Equations (\ref{V1}) (\ref{V-1}) show explicitly how the slope of the
potential can be made arbitrary small in this limit.

Following similar steps
we can integrate the system (\ref{00})--(\ref{kg}) by
quadratures. 
\begin{equation}
\Delta t = {\sqrt{3}\int \frac{da}{a}\left[
{\frac 6{a^6}\int }da{\frac Fa}+\left(C+\frac{A^2}{4}\right){\frac 1{a^6}}+
\frac 3{4a^2}\right]^{-1/2}},
\label{Dt}
\end{equation}
\begin{equation}
\Delta \phi = {\sqrt{6}\int \frac{da}a\left[\frac{-F+6\int da F/a+C}
{6\int da F/a+C+A^2/4+3a^4/4}\right]}^{1/2}.
\label{Dphi}
\end{equation}

In this case, the action of the symmetry transformation
(\ref{tabf})--(\ref{tpbf}) becomes the map $(F(a),C,A)\to(\bar F(\bar a),\bar
C,\bar A)$ where Eq. (\ref{Dt}) remains invariant.  In the case of the
power--law transformation (\ref{tphi})--(\ref{tpo}) we get

\begin{equation} \label{barF}
\bar F=na^{6(n-1)}\left[F+\left(n-1\right)\left(6\int\frac{da}{a}F+
C+\frac{A^2}{4}\right)\right]+
\frac{n\left(3n-1\right)}{4}a^{6n-2}-\frac{a^{4n}}{2}
\end{equation}
\noindent and for models with power--law history of their potential it becomes
\begin{equation} \label{barFp}
\bar F=nB\left(1+6\frac{n-1}{m}\right)a^{6(n-1)+m}+
\frac{n(3n-1)}{4}a^{6n-2}-\frac{a^{4n}}{2}.
\end{equation}

Comparing Eqs. (\ref{Dt}) and (\ref{Dphi}) with Eqs. (\ref{12}) (\ref{13}) we
see that the solutions of FRW for $k=-1$ give the leading behavior of the
solutions of Bianchi V spacetime. Hence, the simple action of the power law
transformation in FRW still holds for Bianchi V in the asymptotic sense.

We recover the solution (\ref{vdot0})--(\ref{vdot1}) inserting

\begin{equation} \label{Fex}
F=\frac{2(2-\omega_0)}{3\omega_0^2}a^{3(2-\omega_0)}-\frac{a^4}{2}
\end{equation}

\noindent into Eqs. (\ref{Va}), (\ref{Dt}) and (\ref{Dphi}), and similarly we
recover the asymptotic behaviour (\ref{poa}) inserting into Eq. (\ref{Dphi})
the leading term of (\ref{Fex}), transformed by (\ref{barF}), when
$\omega_0<2n/3$ .

\section{Conclusions}

We have shown the symmetry transformation under which the Einstein equations
in the general Friedmann--Robertson--Walker  and Bianchi V cosmologies are
form invariant. It relates geometrical quantities with the energy density and
pressure of the perfect fluid. We have seen that the cooperative effect of
adding energy density into the spacetime leads to inflation and gives a way to
link a non-accelerated scenario with an inflationary scenario by means of a
symmetry transformation.

As an example, we have investigated the connections between assisted inflation
and this symmetry transformation. We have shown that assisted inflation can be
generalized to any potential that increases the energy density of the scalar
field configuration without specifying the number of fields.

The violation of the strong energy condition has been shown to be required for
the asymptotic stability of power-law solutions in Bianchi V spacetime, and
the potential leading to these solutions has been found.

We have also constructed the invariant set of perfect fluid
Friedmann--Robertson--Walker and Bianchi V spacetimes that may be obtained by
quadratures. These solutions are characterized by the history of the scalar
field potential and their parametric expression is shown. We give the action
of the symmetry transformation in this representation.

Finally, we conclude that it is very interesting to study this kind of
symmetry transformations, which have received up to now little attention. We
shall continue exploring this subject for other metrics in future papers.


\begin{acknowledgments}

This work was supported by the University of Buenos Aires under Project
X223,  the Spanish Ministry of Science and Technology
jointly with FEDER funds through research grant  BFM2001-0988,
and the University of the Basque Country through research grant 
UPV172.310G02/99. Ruth Lazkoz's
work is also supported by the Basque Government through fellowship BFI01.412.

\end{acknowledgments}

\newpage


\end{document}